\documentclass[preprint,aps]{revtex4}
\usepackage{amsfonts}% Physical Review B

\usepackage{epsf}
\usepackage{anysize}
\marginsize{1 in}{1 in}{1in}{1.40in}

\usepackage{graphicx}% Include figure files
\usepackage{dcolumn}% Align table columns on decimal point
\usepackage{bm}% bold math
\usepackage{amsmath}
\usepackage{amssymb}
\usepackage{mathrsfs}
\usepackage[latin1]{inputenc}
\usepackage{pstricks,pst-node,pst-text,pst-3d}
\usepackage{amsmath}
\usepackage{txfonts}
\usepackage[all]{xy}

\makeatletter
  \newcommand\figcaption{\def\@captype{figure}\caption}
  \newcommand\tabcaption{\def\@captype{table}\caption}
\makeatother

\newcommand{\p}{\partial}
\newcommand{\dd}{{\rm d}}

\newcommand{\e}{\epsilon}

\newcommand{\sd}{Schr\"{o}dinger }
\newcommand{\lmd}{\lambda }
\newcommand{\tr}{{\rm Tr}}

\newcommand{\T}{\mathcal{T}}
\newcommand{\U}{\mathcal{U}}

\newcommand{\hil}{\mathcal{H}}
\newcommand{\h}{{\rm\bf{H}}}

\newcommand{\Q}{{\mathcal{Q}}}

\newtheorem{example}{Example}

\newtheorem{definition}{Definition}

\begin{document}

%\preprint{APS/123-QED}

\title{Singularities of Quantum Control Landscapes}

\author{Rebing Wu}
\altaffiliation{Department of Automation, Tsinghua University,
Beijing, 100084, P.R. China}
%\email{rewu@Princeton.Edu}

\author{Jason Dominy}
\altaffiliation{Program in Applied and Computational Mathematics,
Princeton University, Princeton, New Jersey 08544, USA}

\author{Tak-San Ho}
\altaffiliation{Department of Chemistry, Princeton University,
Princeton, New Jersey 08544, USA}

\author{Herschel Rabitz}
\altaffiliation{Department of Chemistry, Princeton University,
Princeton, New Jersey 08544, USA}

\date{\today}% It is always \today, today,

\begin{abstract}
A quantum control landscape is defined as the objective to be
optimized as a function of the control variables. Existing empirical
and theoretical studies reveal that most realistic quantum control
landscapes are generally devoid of false traps. However, the impact
of singular controls has yet to be investigated, which can arise due
to a singularity on the mapping from the control to the final
quantum state. We provide an explicit characterization of such
controls that are strongly Hamiltonian-dependent and investigate
their associated landscape geometry. Although in principle the
singularities may correspond to local traps, we did not find any in
numerical simulations. Also, as they occupy a small portion of the
entire set of possible critical controls, their influence is
expected to be much smaller than controls corresponding to the
commonly located regular extremals. This observation supports the
established ease of optimal searches to find high-quality controls
in simulations and experiments.
\end{abstract}

%\pacs{03.67.-a,02.20.Qs,02.20.-a  }% PACS, the Physics and Astronomy
%Use  class option if keyword
\maketitle
%\newpage
%\tableofcontents
\section{Introduction}

The concept of a quantum control landscape
\cite{TakRab2006,RabMik2004,RabMik2005,WuRabi2008} was developed to
evaluate the complexity of finding optimal controls, especially with
respect to understanding the observed ease of achieving laser
control of quantum systems \cite{Rabitz2000,BonMit2005}. The search
for optimal controls had been expected to be extremely difficult due
to the complexity of quantum dynamics phenomena. However, practical
studies show that such searches converge rapidly to high-quality
solutions in optimal control theory (OCT) simulations as well as
physically acceptable solutions to optimal control experiments (OCE)
where many additional factors can be involved.

A quantum control landscape \cite{RabMik2004,RabMik2005,WuRabi2008}
is defined as the objective:
\begin{equation}\label{J}
J[\e(\cdot)]=F(\psi(T))
\end{equation}
at some given final time $T$, which is an implicit function of
control field $\e(t)$ that steers the quantum system state $\psi(t)$
in an $N$-dimensional Hilbert space $\mathcal{H}=\mathbb{C}^N$
through satisfaction of the \sd equation:
\begin{equation}\label{control system}
\frac{\dd }{\dd t}\psi(t)=[H_0+\e(t)H_1]\psi(t),~~~~\psi(0)=\psi_0,
\end{equation}where the Planck's constant has been set to $\hbar=1$.
The free and control Hamiltonians, respectively $H_0$ and $H_1$, are
skew-Hermitian operators on the Hilbert space. The goal is to seek
$\max_{\e(\cdot)}J[\e(\cdot)]$, which entails a search on the
landscape aiming to find at least one control that reaches the
absolute global maximal value of $J$.

The efficiency of searching for optimal controls is largely
determined by the topology of the entire set of landscape critical
points among which reside the ultimate desired optimal controls.
Formally, a critical point corresponds to a control $\e(\cdot)$ that
satisfies
\begin{equation}\label{dJ=0}
    \delta J={\left\langle\nabla F(\psi(T)),\delta
    \psi(T)\right\rangle} \equiv 0,~~~\forall~\delta\e(\cdot),
\end{equation}
where the inner product on $\mathbb{C}^N$ is defined as $\langle
v,w\rangle=Re(v^\dagger w)$. The corresponding state variation
$\delta \psi(T)$ at $t=T$ is implicitly dependent on the control
variation $\delta \e(\cdot)$ around $\e(\cdot)$.

In previous studies, we showed that when (i) the system is
controllable at $t=T$, and (ii) any admissible control is regular
(i.e., the mapping from $\delta \e(\cdot)$ to $\delta \psi(T)$ is
surjective), the landscape is topologically equivalent to that of
\begin{equation}\label{Jpsi}
J[\psi]=F(\psi)
\end{equation}
on the unit sphere of $\hil$. Such landscapes are called kinematic
in the sense that their topology is independent of the dynamics. The
study of several classes of quantum control problems
\cite{RabMik2004,RabMik2005,HoJason2009,WuRabi2008,WuRaj2008},
including observable expectation-value optimization and quantum gate
fidelity optimization, revealed that in principle no traps (i.e.,
local suboptima) exist to impede the search for optimal controls,
thereby providing strong support for the observed ease of finding
globally optimal controls in simulations. In the laboratory,
optimization is also very efficient even for highly complex systems,
but constraints on the controls likely imply that less than the
absolute maximum value of the landscape is actually reached.

In general, satisfaction of the controllability assumption is
reasonable under generic circumstances \cite{RamSal1995,WuTarn2006}
because the Lie algebra rank condition for quantum controllability
\cite{RamSal1995,Boothby1975} is easy to fulfill. {However, the
regularity of admissible controls calls for careful assessment
because singular controls may exist corresponding to some $\nabla
F(\psi(T))\neq 0$ without violating the condition (\ref{dJ=0}). In
such a case, the criticality of a control is caused by its
singularity. Understanding whether or not such controls are locally
maximal (i.e., false traps) is important in obtaining a complete
understanding of quantum control landscapes.}

In optimal control theory, there is evidence that singular controls
may become local optima or even global optimal solutions (e.g., in
time optimal control of rockets \cite{Sero2004}). For quantum
systems, very few related studies seem to exist. Boscain and Charlot
\cite{BosCha2004} proved that singular controls cannot be critical
for a class of quantum systems that have multiple independent
control interactions under the rotating wave approximation ;
D'Alessandro showed that at most one singular control can be
critical for the minimal-fluence control of two-level systems and
the control is always constant \cite{Daless2001}. For multi-level
systems, there appears to be no general results.

This paper will give a characterization of singular controls of
single-input quantum systems and investigate their impact on quantum
control landscapes. The balance of this paper is arranged as
follows. Section II defines the singular controls and Section III
provides methods to compute singular controls. In Section IV, the
landscape critical points are classified and computed in numerical
simulations, and their impacts to the landscape are discussed in
Section V. Finally, conclusions are presented in Section VI.

\section{Characterization of singular controls}
From the viewpoint of functional analysis \cite{BonChy2003}, a
control is singular (resp., regular) if the Fr\'echet derivative
\begin{equation}\label{End-point mapping}
    \dd E^{\psi_0,T}:~\delta\e(\cdot)\mapsto \delta\psi(T),
\end{equation} of the end-point mapping $E^{\psi_0,T}:~\e(\cdot)\mapsto \psi(T)$ (defined on a neighborhood of $\e(\cdot)$
for the $L^\infty[0,T]$ norm) is rank deficient (resp., surjective)
from the tangent space of $\e(\cdot)$ to the tangent space
$\T_{\psi(T)}$ of the unit sphere of $\hil$ at $\psi(T)$. The
corresponding $\psi(\cdot)$ is called a singular (resp., regular)
trajectory of (\ref{control system}). It may be shown
\cite{Dominy2008} that $E^{\psi_{0},T}$ is Fr\'echet differentiable
with respect to the $L^{2}$ topology on $[0,T]$ (and therefore also
with respect to the $L^{\infty}$ topology. As a consequence, we can
perturb (\ref{control system}) along the reference trajectory
(driven by the reference control $\e(\cdot)$) to obtain an explicit
form of the derivative:
$$\frac{\dd }{\dd
t}(\psi(t)+\delta \psi(t))=\Big\{H_0+[\e(t)+\delta
\e(t)]H_1\Big\}(\psi(t)+\delta \psi(t)),$$ and then reduce it into
the following time-dependent linear system after omitting
higher-order terms
\begin{equation}\label{linearized control system}
\frac{\dd }{\dd t}\delta \psi(t)=A(t) \delta\psi(t)+B(t)\delta
\e(t),~~~~\delta\psi(0)=0,
\end{equation} where $A(t)=H_0+\e(t)H_1$ and
$B(t)=H_1\psi(t)$. Let $U(t)$ be the system propagator that evolves
$\psi(0)$ to $\psi(t)=U(t)\psi(0)$, then integrating
(\ref{linearized control system}) gives an expression for the
derivative:
\begin{equation}\label{Frechet derivative for quantum}
    \delta \psi(T)=U(T)\int_0^{T}U^{\dagger}(t)H_1\psi(t)\delta \e(t)\dd t=U(T)\int_0^{T}H_1(t)\psi_0\delta \e(t)\dd t,
\end{equation}
where $H_1(t)=U^\dagger(t)H_1U(t)$. Let $M$ be the number of
linearly independent functions of time among the real and imaginary
vector components of $H_1(t)\psi_0$. By the above expression, a
control is regular if there are $M=2N-1$ (i.e., the dimension of
unit sphere in $\hil$) linearly independent functions over $[0,T]$.
Otherwise, the control is singular and the number $k=2N-1-M$ is
called its corank.

In quantum optimal control theory, the cost function often appears
in terms of the system propagator, i.e., $J(\e(\cdot))=F(U(T))$
(e.g., $F(U)=|\tr(W^\dag U)|$ for maximizing the gate fidelity with
some specified unitary $W$), where $U(t)$ obeys the evolution
equation
\begin{equation}\label{control system U}
\frac{\dd }{\dd t}U(t)=[H_0+\e(t)H_1]U(t),~~~~U(0)=I_N.
\end{equation}
{For such problems, the singular controls correspond to those such
that the Fr\'echet derivative
\begin{equation}\label{End-point mapping}
    \dd E^{\psi_0,T}:~\delta\e(\cdot)\mapsto \delta U(T),
\end{equation} of the end-point mapping $E^{\psi_0,T}:~\e(\cdot)\mapsto U(T)$ (defined on a neighborhood of $\e(\cdot)$
for the $L^\infty[0,T]$ norm) is rank deficient from the tangent
space of $\e(\cdot)$ to the tangent space $\T_{U(T)}$ of the unitary
group $\U(N)$ at $U(T)$. One can similarly derive the Frechet
derivative from $\delta \e( \cdot)$ to $\delta U(T)$ at $U(T)$
\begin{equation}\label{Frechet derivative for U}
    \delta U(T)=U(T)\int_0^{T}H_1(t)\delta \e(t)\dd
    t,
\end{equation}
according to which a control is singular to the
control-to-propagator mapping if $H_1(t)$ contains less than $N^2$
linearly independent functions.} For example:
\begin{example}
Any constant control is singular to the control-to-propagator
mapping. Their coranks are at least $N-1$.
\end{example}
\textbf{Proof:} Let $\e(t)\equiv c$ be a constant control, then its
singularity is equivalent to the linear dependence of the matrix
elements of $$H_1(t)=\exp(-t(H_0+cH_1))H_1\exp(t(H_0+cH_1))$$ as
functions of time. Suppose that $H_0+cH_1=Q\Lambda Q^\dag$ where
$\Lambda$ is diagonal and $Q$ is a unitary transformation, then the
analysis is equivalent to investigating the matrix elements of
$$\tilde H_1(t)=Q^\dag H_1(t)Q=\exp(-t\Lambda)\tilde H_1\exp(t\Lambda),\quad \tilde H_1=Q^\dag H_1Q.$$
The $N$ diagonal matrix elements of $\tilde H_1(t)$ are all
constant, implying that they are mutually linearly dependent,
thereby the corank is at least $N-1$. The corank increases when
$\tilde H_1$ has fewer nonzero eigenvalues or $H_0+cH_1$ has a
degenerate spectrum. Q.E.D.

{It should be noted that the condition for a control to be singular
to the control-to-state mapping is stronger than that to the
control-to-propagator mapping, i.e., any singular control (e.g., the
constant control) for the control-to-state mapping must also be
singular for the control-to-propagator mapping, as can be easily
seen from (\ref{Frechet derivative for quantum}) and (\ref{Frechet
derivative for U}), but the inverse is not true.} There exist
singular controls for the control-to-propagator mapping that are not
singular for the control-to-state mapping.

\section{Computation of singular controls}

This section will provide two approaches to numerically compute
singular controls from different perspectives. The first one derives
the singular controls by projecting the singular trajectories from a
lifted space. The second one directly gives the control in a
feedback form, however, additional smoothness constraints on the
control are posed.

Firstly, according to (\ref{dJ=0}) and (\ref{Frechet derivative for
quantum}), a control $\e(\cdot)$ is singular if and only if there
exists a nonzero vector $\phi_T\in\mathcal{T}_{\psi(T)}$ such that
\begin{equation}\label{dU/de=0}
\left\langle\phi_T,
U(T)U^{\dagger}(t)H_1\psi(t)\right\rangle=\left\langle
U(t)U^{\dagger}(T)\phi_T, H_1\psi(t)\right\rangle=0,\quad \forall~
t\in[0,T].
\end{equation}By defining the conjugate vector
$\phi(t)=U(t)U^{\dagger}(T)\phi_T\in\mathcal{T}_{\psi(t)}$, then we
have
\begin{eqnarray}
\frac{\dd }{\dd
t}\psi(t)&=&[H_0+\e(t)H_1]\psi(t),~~~~\psi(0)=\psi_0,
 \label{PMP1}\\
\frac{\dd }{\dd
t}\phi(t)&=&[H_0+\e(t)H_1]\phi(t),~~~~\phi(T)=\phi_T, \label{PMP2}
\end{eqnarray}subject to the algebraic constraint from (\ref{dU/de=0})
\begin{equation}\label{H1t} \langle\phi(t),H_1\psi(t)\rangle\equiv0,\quad
\forall~t\in[0,T].
\end{equation}This is a two-point boundary-value problem in time
which has to be solved by iterative numerical algorithms. As will be
seen later, these equations can be also derived from the Pontryagin
Maximum Principle with respect to a given cost function, with
$\phi(T)$ assigned to be the gradient vector of the cost function.

Using (\ref{PMP1}) and (\ref{PMP2}), the equation (\ref{H1t}) can be
differentiated to derive a new algebraic constraint:
\begin{equation}\label{H01(t)=0}
\frac{\dd}{\dd t}\langle \phi(t),H_1\psi(t)\rangle\equiv
0~~\Rightarrow~~\left\langle
\phi(t),[H_0,H_1]\psi(t)\right\rangle\equiv0,
\end{equation}
which can be again differentiated to arrive at an explicit
relationship between a singular control $\e(\cdot)$ and the
corresponding state trajectory:
\begin{equation}\label{d2H}
\frac{\dd^2}{\dd
t^2}\langle\phi(t),H_1\psi(t)\rangle\equiv0~~\Rightarrow~~\langle\phi(t),[H_0,[H_0,H_1]]\psi(t)\rangle
+\e(t)\langle\phi(t),[H_1,[H_0,H_1]]\psi(t)\rangle=0.
\end{equation}

According to (\ref{d2H}), we classify singular controls as
follows:

(i) When $\langle \phi(t),[H_1,[H_1,H_0]]\psi(t)\rangle\neq 0$,
$\forall t\in[0,T]$, the singular control can be expressed in a
feedback form
\begin{equation}\label{singular control psi}
\e(t)=-\frac{\langle\phi(t),[H_0,[H_0,H_1]]\psi(t)\rangle}{\langle\phi(t),[H_1,[H_0,H_1]]\psi(t)\rangle}.
\end{equation}
One can then combine (\ref{PMP1}), (\ref{PMP2}) and (\ref{singular
control psi}) to solve for the singular control. Such controls are
called minimal-order singular controls. Notice that since each
nonzero $\phi_T$ must uniquely correspond to some
$\phi(0)=\phi_0\neq 0$ at $t=0$, we can equivalently integrate the
differential equations (\ref{PMP1}) and (\ref{PMP2}) from $t=0$
(this can be done from the other end $t=T$ as well) for any given
pair of $(\psi_0,\phi_0)$ that satisfy
$$\langle\phi_0,H_1\psi_0\rangle=\langle\phi_0,[H_0,H_1]\psi_0\rangle=0,\quad\langle\phi_0,[H_1,[H_0,H_1]]\psi_0\rangle\neq0.$$
Moreover, since (\ref{PMP1}) and (\ref{PMP2}) share the same
evolution propagator $U(t)$, we can use the dynamics of the system
propagator:
\begin{equation}\label{STpsi} \frac{\dd U(t)}{\dd t} = \left(H_0-\frac{\langle\phi_0,U^\dag(t)[H_0,[H_0,H_1]]U(t)\psi_0\rangle}
{\langle\phi_0,U^\dag(t)[H_1,[H_0,H_1]]U(t)\psi_0\rangle}
H_1\right)U(t), \quad U(0)=I_N,
\end{equation}to obtain a singular trajectory in the unitary group $\U(N)$
parameterized by $(\psi_0,\phi_0)$. In this way, singular controls
can be systematically generated  without iterative computations.

(ii) When $\langle \phi(t),[H_1,[H_1,H_0]]\psi(t)\rangle=0$ on
$[0,T]$, this implies $\langle
\phi(t),[H_0,[H_1,H_0]]\psi(t)\rangle=0$ as well. In such cases,
equation (\ref{singular control psi}) is not sufficient for
determining a singular control. However, one may go on
differentiating these two quantities until $\e(t)$ can be explicitly
expressed. Let $H_{\alpha_1\cdots
\alpha_k}=[H_{\alpha_1},[H_{\alpha_2},[\cdots,[H_{\alpha_{k-1}},H_{\alpha_k}]\cdots]]]$.
If there exists a finite integer $k\geq 2$ such that
$\langle\phi(t),H_{\beta_1\cdots \beta_\ell}\psi(t)\rangle\equiv0$
for any $(\beta_1,\ldots,\beta_\ell)\in\{0,1\}^\ell$, $2\leq
\ell\leq k$, but the differentiation of some
$\langle\phi(t),H_{\alpha_1\ldots\alpha_k}\psi(t)\rangle$ gives
\begin{equation}\label{dkH}
\langle\phi(t),H_{0\alpha_1\cdots \alpha_k}\psi(t)\rangle
+\e(t)\langle\phi(t),H_{1\alpha_1\cdots \alpha_k}\psi(t)\rangle=0,
\end{equation}where $\langle\phi(t),H_{1\alpha_1\cdots
\alpha_k}\psi(t)\rangle\neq 0$, then a singular control can be
formally obtained from a feedback equation
\begin{equation}\label{STpsi-k} \frac{\dd U(t)}{\dd t} = \left(H_0-\frac{\langle\phi_0,U^\dag(t)H_{0\alpha_1\cdots \alpha_k}U(t)\psi_0\rangle}
{\langle\phi_0,U^\dag(t)H_{1\alpha_1\cdots
\alpha_k}U(t)\psi_0\rangle} H_1\right)U(t), \quad U(0)=I_N,
\end{equation}
and we call it a $k$-th order singular control.

Let $\mathcal{B}^{(1)}={\rm span}\{H_1\}$ and
$\mathcal{B}^{(\ell)}={\rm span}\{H_{\alpha_1\cdots
\alpha_\ell}|\alpha_1,\ldots,\alpha_\ell=0,1\}$ ($\ell\geq2$) be the
subspaces of skew-Hermitian matrices generated by $\ell$ tuples of
commutations. A geometrical interpretation for a control to be
$k$-th order singular is that the adjoint vector
$\phi(t)\in\mathcal{T}_{\psi(t)}$ is orthogonal to
$\mathcal{B}^{(\ell)}\psi(t)\subset \mathcal{T}_{\psi(t)}$ for any
$1\leq \ell\leq k$ but not orthogonal to
$\mathcal{B}^{(k+1)}\psi(t)$, i.e., $\phi(t)$ belongs to the
following $k$th order singular cone at $\psi(t)$:
$$\mathcal{V}^{(k)}_{\psi(t)}=\left\{\phi\in \mathcal{T}_{\psi(t)}~\Big|~\left\langle\phi, \bigcup_{\ell=1}^k\mathcal{B}^{(\ell)}\psi(t)\right\rangle=0,~ \langle\phi, \mathcal{B}^{(k+1)}\psi(t)\rangle\neq0\right\}.$$
To locate a $k$-th order singular control, one can choose a pair
$(\psi_0,\phi_0)$ such that $\phi_0\in \mathcal{V}^{(k)}_{\psi_0}$,
and integrate (\ref{STpsi-k}) from $t=0$, provided the solution
exists and is unique.

(iii) When the control function does not explicitly appear in
(\ref{dkH}) for any integer $k\in\mathbb{N}$, then its order is
infinite. Let $\mathcal{L}$ be the Lie algebra generated by $H_0$
and $H_1$, and
$\mathcal{L}_0=\bigcup_{\ell=1}^\infty\mathcal{B}^{(\ell)}$ be the
minimal ideal in $\mathcal{L}$ that contains $H_1$. The codimension
of $\mathcal{L}_0\psi_0$ in $\mathcal{L}$ is either 0 or 1. For
infinite-order singular controls, the codimension of
$\mathcal{L}_0\psi_0$ in $\mathcal{L}\psi_0$ must be 1 (otherwise
$\phi(t)$ has to vanish), i.e., the adjoint vector
$\phi(t)\in\mathcal{T}_{\psi(t)}$ varies in the one-dimension
complimentary subspace of $\mathcal{L}_0\psi(t)$ in
$\mathcal{L}\psi(t)$.

It is also possible for the denominator in (\ref{singular control
psi}) to cross zero at isolated time instants, which divide a
singular control into pieces of singular ``arcs" {whose orders may
be different with each other}. In this paper, we only consider
singular controls whose order is constant on $[0,T]$.

An alternative approach to produce (i) and (ii) above is as follows.
Denote the $2N-1$ independent elements in the vector $H_1(t)\psi_0$
by $\xi(t)=(\xi_1(t),\cdots,\xi_{2N-1}(t))$. For a singular control,
there must exist a nonzero constant vector
$\phi_0=(c_1,\cdots,c_{2N-1})$ such that
$\sum_{i=1}^{2N-1}c_i\xi_i(t)\equiv 0$, $\forall~t\in[0,T]$, which
can be repeatedly differentiated to give
$$\sum_{i=1}^{2N-1}c_i\xi_i^{(k)}(t)\equiv 0,\quad k=1,\cdots,2N-2.$$
So we have
$$\left(            \begin{array}{cccc}
                      \xi_1(t) & \xi_2(t) &\cdots&\xi_{2N-1}(t) \\
                      \xi_1^{(1)}(t) & \xi_2^{(1)}(t) &\cdots&\xi_{2N-1}^{(1)}(t) \\
                      \vdots & \vdots & \ddots & \vdots \\
                      \xi_1^{(2N-2)}(t) & \xi_2^{(2N-2)}(t) &\cdots&\xi_{2N-1}^{(2N-2)}(t) \\
                    \end{array}
                  \right)\left(
                           \begin{array}{c}
                             c_1 \\
                             c_2 \\
                             \vdots \\
                             c_{2N-1} \\
                           \end{array}
                         \right)\equiv 0,\quad \forall t\in[0,T]
$$which implies that the Wronskian must vanish, i.e.,
$$\left|
   \begin{array}{cccc}
     \xi_1(t) & \xi_2(t) &\cdots&\xi_{2N-1}(t) \\
     \xi_1^{(1)}(t) & \xi_2^{(1)}(t) &\cdots&\xi_{2N-1}^{(1)}(t) \\
     \vdots & \vdots & \ddots & \vdots \\
     \xi_1^{(2N-2)}(t) & \xi_2^{(2N-2)}(t) &\cdots&\xi_{2N-1}^{(2N-2)}(t) \\
   \end{array}
 \right|\equiv0.$$
From the previous derivations, $\xi_i^{(2)}(t)$ can be decomposed
into two parts as $\xi_i^{(2)}(t)=a_i^{(2)}(t)+\e(t)b_i^{(2)}(t)$,
where $a_i(t)$ and $b_i(t)$ correspond to linearly independent
functions in $[H_0,[H_0,H_1]](t)\psi_0$ and
$[H_1,[H_0,H_1]](t)\psi_0$, respectively
\begin{equation}\label{wronskian}
\left|
   \begin{array}{cccc}
     \xi_1(t) & \xi_2(t) &\cdots&\xi_{2N-1}(t) \\
     \xi_1^{(1)}(t) & \xi_2^{(1)}(t) &\cdots&\xi_{2N-1}^{(1)}(t) \\
     a_1^{(2)}(t) & a_2^{(2)}(t) &\cdots& a_{2N-1}^{(2)}(t) \\
     \vdots & \vdots & \ddots & \vdots \\
     \xi_1^{(2N-2)}(t) & \xi_2^{(2N-2)}(t) &\cdots&\xi_{2N-1}^{(2N-2)}(t) \\
   \end{array}
 \right|+\e(t)\left|
   \begin{array}{cccc}
     \xi_1(t) & \xi_2(t) &\cdots&\xi_{2N-1}(t) \\
     \xi_1^{(1)}(t) & \xi_2^{(1)}(t) &\cdots&\xi_{2N-1}^{(1)}(t) \\
     b_1^{(2)}(t) & b_2^{(2)}(t) &\cdots& b_{2N-1}^{(2)}(t) \\
     \vdots & \vdots & \ddots & \vdots \\
     \xi_1^{(2N-2)}(t) & \xi_2^{(2N-2)}(t) &\cdots&\xi_{2N-1}^{(2N-2)}(t) \\
   \end{array}
 \right|\equiv0,\end{equation}
{where the time derivatives of $\e(t)$ (up to the $(2N-4)$-th order)
are involved in the fourth to the $(2N-2)$-th rows. Hence
(\ref{wronskian}) forms an ordinary differential equation (up to
$(2N-4)$-th order) of $\e(t)$ whose coefficients are functions of
the system propagator $U(t)$. This relation can be computed in the
\sd equation to solve for the singular controls from given initial
values of $\e(t)$ and its derivatives.}

{Both the above two approaches calculate singular controls via
ordinary differential equations from the same condition
(\ref{Frechet derivative for quantum}). The solution by the first
approach is parameterized by a prescribed vector $\phi_0$ as the
initial condition of the conjugate equation, while the latter is by
the initial conditions of the time derivatives of $\e(t)$. Moreover,
it is easy to see that the order of the differential equation
(\ref{wronskian}) with respect to $\e(t)$ is $2N-2-k$, as the
derivatives of $\e(t)$ start to appear from the $(k+2)$-th row,
where $k$ is the order of the singular control defined in the first
approach. In this regard, these two approaches are equivalent. In
comparison, the first approach is numerically more efficient and
will be adopted in the simulation examples below. The latter is
conceptually simple because it appears directly as {the sum of the
derivatives of the control function and the state} without
introduction of any conjugate vector, and hence provides a useful
perspective for the origin of singularity. }

\section{Singular controls as
landscape critical points} {As analyzed above, the critical points
for a given control landscape can be a regular or singular control.
The corresponding kinematic gradient must vanish when the critical
point is a regular control , but it may not vanish when the critical
point is a singular control.} In this regard, we can classify the
landscape critical points into the following three categories:
\begin{definition}
A control is said to be regularly (singularly) kinematic if it is
regular (singular) and $\nabla F(\psi(T))=0$. Otherwise, if $\nabla
F(\psi(T))\neq 0$, it is said to be non-kinematic.
\end{definition}

Let $\mathcal{C}_F=\{\psi(T)\in\mathcal{S}_\hil~|~\nabla
F(\psi(T))=0\}$, where $\mathcal{S}_\hil$ is the unit sphere in
$\hil$, be the set of kinematic (either regularly or singularly)
critical points. Any control that steers the trajectory onto
$\mathcal{C}_F$ at $t=T$ must be a kinematic critical control. As
indicated in previous studies \cite{TakRab2006}, the dimension of
regularly kinematic controls is infinite and its codimension in the
set of admissible controls is the same as that of $\mathcal{C}_F$ in
$\hil$.

For nonkinematic critical points, their corresponding kinematic
gradient vector $\phi_T=\nabla F(\psi(T))$ must be nonzero and
belong to some singular cone $\mathcal{V}^{(k)}_{\psi(T)}$ ($2\leq
k\leq \infty$). This criterion forms a nonlinear constraint on the
final state $\psi(T)$ and thereby defines a subset of $\hil$:
\begin{equation}\label{J surface}
\mathcal{S}^{(k)}_F=\Big\{\psi\in\mathcal{S}_\hil~\Big|~\nabla
F(\psi)\in \mathcal{V}^{(k)}_{\psi}\Big\},
\end{equation}
which will be called the $k$-th order singular surface for the
control landscape $F$. Every $k$-th order singular control has to
terminate at this surface to become a nonkinematic critical point.
Starting from an arbitrary point $\psi_T\in\mathcal{S}^{(k)}_F$, one
can determine a constant $k$-th order nonkinematic critical point by
integrating (\ref{PMP1}) and (\ref{PMP2}) backwards in time along
any proper direction in $\mathcal{V}^{(k)}_{\psi_T}$ if the solution
exists.

\begin{figure}[h]
\centerline{
\includegraphics[width=4in,height=3in]{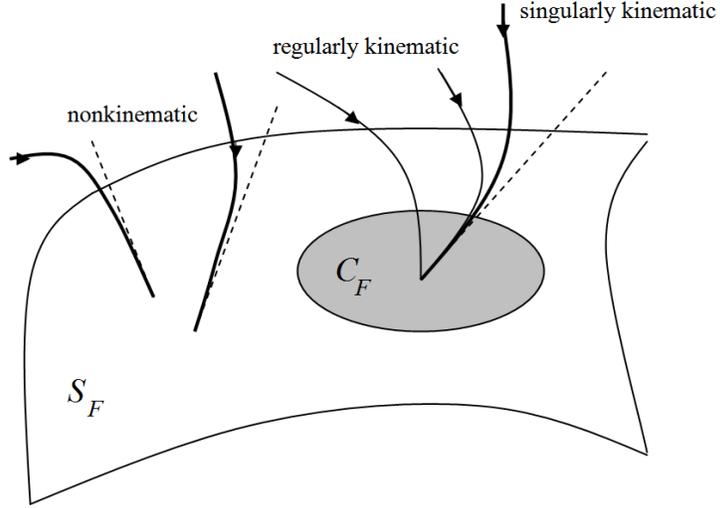}
} \caption{Schematic illustration for the geometry of the set of
extremal controls. Every extremal trajectory ends up at the singular
surface $\mathcal{S}_F$ at $t=T$. The singular trajectories (bold
lines) follow the flow defined by (\ref{STpsi}) and traverse the
singular surface along some direction (dash lines) in the singular
cone, among which those that traverse $\mathcal{C}_F$ are singularly
kinematic, and the remainder are nonkinematic. Any other control
that steers the trajectory to traverse $\mathcal{C}_F$ is regularly
kinematic.}\label{SingularSurface}
\end{figure}

It is difficult to estimate the dimension of the entire set of
nonkinematic controls, as they can be a combination of singular arcs
with different orders, which correspond to an infinite number of
possibilities. Nonetheless, as illustrated in Fig.
\ref{SingularSurface}, the set of regularly kinematic critical
controls is much richer than nonkinematic ones because they may
cross the surface $\mathcal{C}_F$ along any direction and with any
admissible flows. By contrast, the set of singular (kinematic and
nonkinematic) critical points is much more limited.

\section{The role of singularities in quantum control landscapes}
{We have derived that all of the nonkinematic critical points are
subject to the equations (\ref{PMP1}) and (\ref{PMP2}) with the
algebraic constraint $\langle\phi(t),H_1\psi(t)\rangle=0$, where the
boundary condition satisfies $\phi_T=\nabla F(\psi(T))\neq 0$. On
the other hand, any (regularly or singularly) kinematic critical
points automatically satisfy the same differential equations and the
algebraic constraint because $\phi_T=\nabla F(\psi(T))=0$.
Therefore, in the language of the Pontryagin Maximum Principle
\cite{Stengel1994}, we can unify the conditions for a control to be
a landscape critical point as:}
\begin{equation}\label{PMP}\frac{\dd \psi}{\dd t}=\frac{\p \h}{\p
\phi},\quad \frac{\dd \phi}{\dd t}=-\frac{\p \h}{\p \psi},\quad
\frac{\p\h}{\p\e}\equiv0, \quad \psi(0)=\psi_0, \quad \phi_T=\nabla
F(\psi(T)), \end{equation} where $\h[\e(t),\psi(t),\phi(t)]=\langle
\phi(t),H_0\psi(t)\rangle+\e(t)\langle \phi(t),H_1\psi(t)\rangle$ is
the pseudo-Hamiltonian function.

In common optimal control problems, the dynamics are often taken
into account through the cost function
\begin{equation}\label{J-L}
J[\e(\cdot)]=F(\psi(T))+\int_0^TL(\psi(t),\e(t),t)\dd t,
\end{equation} where $L(\cdot)$ is a function chosen to balance
the dynamical performance issues. The standard Pontryagin maximum
principle then corresponds to the following pseudo-Hamiltonian
function to solve for critical points of $J$ (conventionally called
extremals):
$$\h[\psi(t),\phi(t),\e(t)]=\lmd L(\psi(t),\e(t),t)+\langle \phi(t),H_0\psi(t)\rangle+\e(t) \langle \phi(t),H_1\psi(t)\rangle,~~\phi(T)=\nabla F(\psi(T)),$$
where $\lmd$ is a constant. Taking the optimization process as a
dynamical game between the end-point cost (the first term) and the
dynamical part (the second term), {the extremals corresponding to
$\lmd\neq0$ result from the trade-off between the two costs, under
which the system cannot attain perfect yields\cite{ZhuRab1998}
(i.e., the highest yield in the control landscape (\ref{J})). Such
extremals are conventionally called normal extremals, which are
generally not critical points of (\ref{J}). Normal extremals are
always regular because the necessary conditions (\ref{PMP1}) and
(\ref{PMP2}) are not satisfied.}

\begin{figure}[h]
\centerline{
\includegraphics[width=4.5in,height=1.3in]{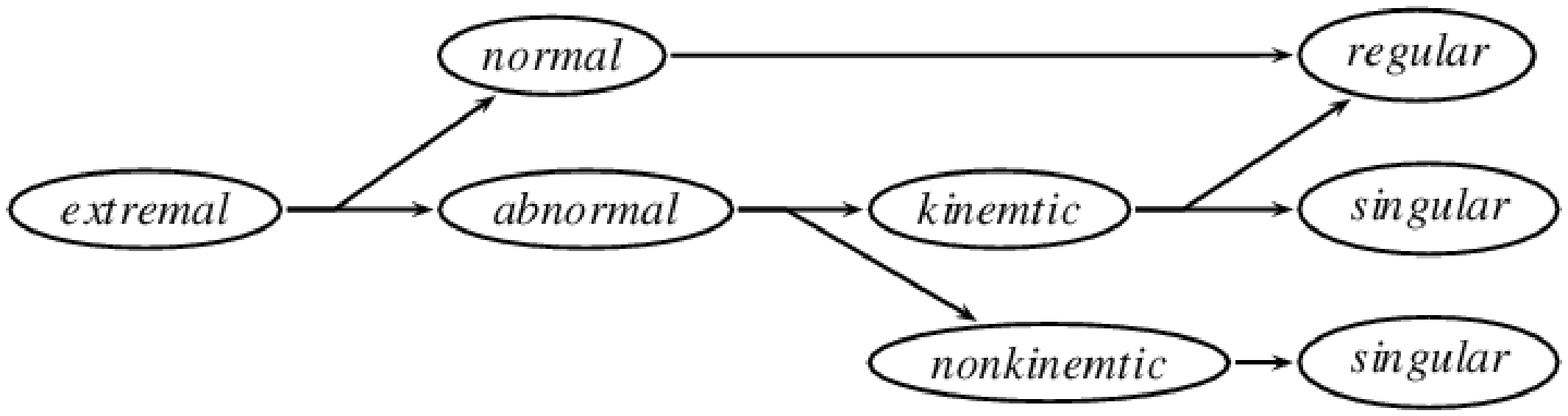}
} \caption{Classification of extremal controls for general optimal
control problems.}\label{classification}
\end{figure}

Extremals corresponding to $\lmd=0$ are called abnormal, and they
are also critical points of (\ref{J}). Under such controls, the
end-point cost completely overwhelms the dynamical cost so that the
resulting trajectories are independent of the choice of the cost
function $L(\cdot)$ in the integral. They include all kinematic and
non-kinematic critical points discussed in this paper (there is no
analog of the kinematic picture for normal extremals because the
dynamics is always relevant). Thus, we can classify the extremal
controls as in Fig. \ref{classification}.

The optimality of abnormal extremals can be analyzed through the
second-order variation of (\ref{J}):
\begin{equation}\label{d2J}
  \delta^2J =  \Q_{\psi(T)}(\delta\psi(T),\delta\psi(T))+\langle\nabla
  J(\psi(T)),\delta^2\psi(T)\rangle,
\end{equation}
where the kinematic part $\Q_{\psi(T)}$ is the Hessian quadratic
form at $\psi(T)$. The second term vanishes when the control is a
kinematic extremal, either regular or singular, and hence leaves the
optimality determined solely by the positive-definiteness of the
kinematic Hessian form $\Q_{\psi(T)}$ on the set of achievable
$\delta\psi(T)$ at ${\psi(T)}$. For regular abnormal extremals
(i.e., regularly kinematic critical points), the optimality is
exactly the same as that of the kinematic Hessian, and hence their
topology is reflected by that of the corresponding final state
$\psi(T)$ in the kinematic picture, exhibiting a universal
Hamiltonian-independent property.

If an abnormal extremal is a singularly kinematic critical point,
then its Hessian form $\Q_{\psi(T)}$ is defined on the range of the
Frechet derivative $\dd E^{\psi_0,T}$ as a proper subspace of
$\T_{\psi(T)}$. When the unrestricted $\Q_{\psi(T)}$ is positive
(negative) semi-definite, the restricted $\Q_{\psi(T)}$ is also
positive (negative) semi-definite. {However, in the case that the
unrestricted $\Q_{\psi(T)}$ has both positive and negative
eigenvalues, its restriction on the range of $\dd E^{\psi_0,T}$ may
be positive (negative) when the range of $\dd E^{\psi_0,T}$ is
contained in the subspace spanned by the eigenvectors corresponding
to positive (negative) eigenvalues. This implies that a saddle point
may degenerate to a local minimum (maximum).}

\begin{figure}[h]
\centerline{
\includegraphics[width=6in,height=2.5in]{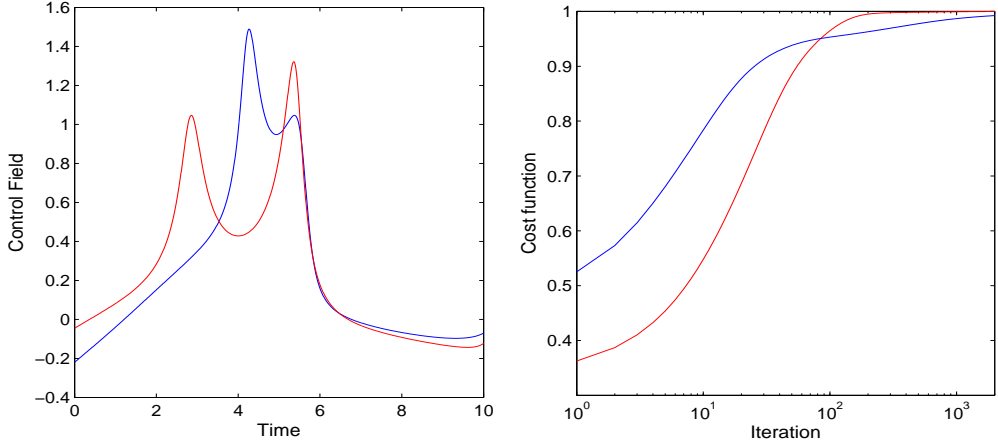}
} \caption{Nonkinematic critical points of the quantum state
transition control landscape
$J[\e(\cdot)]=|\psi_f^\dagger\psi(T)|^2$ for a four-level system.
The figure on the left displays two singular extremal controls from
randomly chosen initial states. Starting from some control $\e_{\rm
initial}(t)$ in a small neighborhood ($\|\e_{\rm
initial}(\cdot)-\e(\cdot)\|\leq 0.01$) of the singular control
$\e_0(t)$, the steepest ascent algorithm is performed to search for
optimal controls. Both of them climb and are not attracted by the
corresponding singular controls, but approach a perfect yield,
showing that these singular controls are not local false traps.
}\label{STexample}
\end{figure}

Nonkinematic critical points are beyond the scope of the kinematic
picture, for which the second-order variation $\delta^2\psi(T)$
related to the system dynamics is nonvanishing. Hence, the
optimality of nonkinematic critical points is much more complex to
assess as the Hessian form has to be discussed on an infinite
dimensional space of control fields. As indicated by Bonnard and
Chyba\cite{BonChy2003}, there is a possibility that such critical
points are local optima in the control landscape. However, in our
numerical simulations, no such traps have been found.
Fig.\ref{STexample} shows examples of nonkinematic extremal controls
for the quantum state transition control landscape
$J[\e(\cdot)]=|\psi_f^\dagger\psi(T)|^2$ for a four-level quantum
system, where
$$H_0 =\left(
        \begin{array}{cccc}
          -0.50 &  &  &  \\
           & 0.00 &  &  \\
           &  & 0.20 &  \\
           &  &  & 0.60 \\
        \end{array}
      \right),\quad H_1=\left(
                          \begin{array}{cccc}
                           0.30  & 0.75-0.20i & 0.65 &0.40\\
                            0.75+0.20i & 0.70 &0.70-0.50i & 0.20+0.30i  \\
                            0.65 & 0.70+0.50i & 0.30 &0.50 \\
                            0.40 & 0.20 -0.30i &0.50 & 0.60 \\
                          \end{array}
                        \right),
$$ with the initial and target states being $\psi(0)=[1 ~0~ 0~ 0]^T$ and $\psi_f=[0 ~0~ 0~ 1]^T$, respectively. Singular trajectories
were found by solving equation (\ref{STpsi}). After locating such
singular controls, we started a gradient search from small
neighborhoods of the controls, and found that the search always
climbed towards perfect yield ($J=1.0$) without being trapped. This
behavior is consistent with the observation that singular controls
have not been located when performing common optimal control
simulations, i.e., a gradient flow trajectory is not attracted to a
singular trajectory even when one is nearby.

\section{Conclusion}
This paper considered the role of singular controls upon the search
for optimal solutions over quantum control landscapes. We indicate
that the regularity assumption of admissible controls can be
violated, which gives rise to singularly kinematic or non-kinematic
critical points that are beyond the scope of the kinematic picture.
For single-input systems, these singular controls could possibly be
locally optimal. However, such cases were not found in our
simulations. Moreover, since the entire set of singular controls is
small in contrast with that of the regular controls, the overall
landscape is not expected to be rugged. Hence, regular controls
dominate over the singular ones, and the regular controls should
still determine the overall landscape complexity. {This conclusion
is fully consistent with the general quantum control literature
where no specific evidence  is seen for singular controls to be
local sub-optimal solutions.}

{When the system has multiple control fields associated with
independent operators $H_k$:
\begin{equation}\label{multiple control system}
\frac{\dd }{\dd
t}\psi(t)=\left[H_0+\sum_{k=1}^m\e_k(t)H_k\right]\psi(t),~~~~\psi(0)=\psi_0,
\end{equation}
the definitions of extremals can be extended but will not be given
here. In this case the condition for a control to be singular and
further critical is much more stringent. In particular, Chitour {\it
et al}\cite{Chitour2008} proved a generic property for the system
class represented by $(m+1)$-tuple $(H_0,H_1,\cdots,H_m)$, showing
that almost all such systems do not possess minimizing singular
controls. Therefore, the impact of singular controls on the control
landscape is smaller yet, or even disappears, for multi-control
quantum systems.}

\begin{acknowledgments}
The authors acknowledge support from the NSF. JD acknowledges
support from the Program in Plasma Science and Technology (PPST) at
Princeton. Useful discussions with Professor Bernard Bonnard are
appreciated.
\end{acknowledgments}

%\bibliographystyle{IEEEtr}
%\bibliography{STRef}

\begin{thebibliography}{20}
\expandafter\ifx\csname
natexlab\endcsname\relax\def\natexlab#1{#1}\fi
\expandafter\ifx\csname bibnamefont\endcsname\relax
  \def\bibnamefont#1{#1}\fi
\expandafter\ifx\csname bibfnamefont\endcsname\relax
  \def\bibfnamefont#1{#1}\fi
\expandafter\ifx\csname citenamefont\endcsname\relax
  \def\citenamefont#1{#1}\fi
\expandafter\ifx\csname url\endcsname\relax
  \def\url#1{\texttt{#1}}\fi
\expandafter\ifx\csname urlprefix\endcsname\relax\def\urlprefix{URL
}\fi \providecommand{\bibinfo}[2]{#2}
\providecommand{\eprint}[2][]{\url{#2}}

\bibitem[{\citenamefont{Ho and Rabitz}(2006)}]{TakRab2006}
\bibinfo{author}{\bibfnamefont{T.-S.} \bibnamefont{Ho}} \bibnamefont{and}
  \bibinfo{author}{\bibfnamefont{H.}~\bibnamefont{Rabitz}},
  \bibinfo{journal}{J. Photochemistry Photobiology A}
  \textbf{\bibinfo{volume}{180}}, \bibinfo{pages}{226} (\bibinfo{year}{2006}).

\bibitem[{\citenamefont{Rabitz et~al.}(2004)\citenamefont{Rabitz, Hsieh, and
  Rosenthal}}]{RabMik2004}
\bibinfo{author}{\bibfnamefont{H.}~\bibnamefont{Rabitz}},
  \bibinfo{author}{\bibfnamefont{M.}~\bibnamefont{Hsieh}}, \bibnamefont{and}
  \bibinfo{author}{\bibfnamefont{C.}~\bibnamefont{Rosenthal}},
  \bibinfo{journal}{Science} \textbf{\bibinfo{volume}{303}},
  \bibinfo{pages}{1998} (\bibinfo{year}{2004}).

\bibitem[{\citenamefont{Rabitz et~al.}(2005)\citenamefont{Rabitz, Hsieh, and
  Rosenthal}}]{RabMik2005}
\bibinfo{author}{\bibfnamefont{H.}~\bibnamefont{Rabitz}},
  \bibinfo{author}{\bibfnamefont{M.}~\bibnamefont{Hsieh}}, \bibnamefont{and}
  \bibinfo{author}{\bibfnamefont{C.}~\bibnamefont{Rosenthal}},
  \bibinfo{journal}{Phys. Rev. A} \textbf{\bibinfo{volume}{72}},
  \bibinfo{pages}{52337} (\bibinfo{year}{2005}).

\bibitem[{\citenamefont{Wu et~al.}(2008{\natexlab{a}})\citenamefont{Wu, Rabitz,
  and Hsieh}}]{WuRabi2008}
\bibinfo{author}{\bibfnamefont{R.}~\bibnamefont{Wu}},
  \bibinfo{author}{\bibfnamefont{H.}~\bibnamefont{Rabitz}}, \bibnamefont{and}
  \bibinfo{author}{\bibfnamefont{M.}~\bibnamefont{Hsieh}}, \bibinfo{journal}{J.
  Phys. A} \textbf{\bibinfo{volume}{41}}, \bibinfo{pages}{015006}
  (\bibinfo{year}{2008}{\natexlab{a}}).

\bibitem[{\citenamefont{Rabitz et~al.}(2000)\citenamefont{Rabitz,
  de~Vivie-Riedle, Motzkus, and Kompa}}]{Rabitz2000}
\bibinfo{author}{\bibfnamefont{H.}~\bibnamefont{Rabitz}},
  \bibinfo{author}{\bibfnamefont{R.}~\bibnamefont{de~Vivie-Riedle}},
  \bibinfo{author}{\bibfnamefont{M.}~\bibnamefont{Motzkus}}, \bibnamefont{and}
  \bibinfo{author}{\bibfnamefont{K.}~\bibnamefont{Kompa}},
  \bibinfo{journal}{Science} \textbf{\bibinfo{volume}{288}},
  \bibinfo{pages}{824} (\bibinfo{year}{2000}).


\bibitem[{\citenamefont{Bonacic-Koutechy and Mitric}(2005)}]{BonMit2005}
\bibinfo{author}{\bibfnamefont{V.}~\bibnamefont{Bonacic-Koutechy}}
  \bibnamefont{and} \bibinfo{author}{\bibfnamefont{R.}~\bibnamefont{Mitric}},
  \bibinfo{journal}{Chem. Rev.} \textbf{\bibinfo{volume}{105}},
  \bibinfo{pages}{11} (\bibinfo{year}{2005}).

\bibitem[{\citenamefont{Ho et~al.}(2009)\citenamefont{Ho, Dominy, and
  Rabitz}}]{HoJason2009}
\bibinfo{author}{\bibfnamefont{T.-S.} \bibnamefont{Ho}},
  \bibinfo{author}{\bibfnamefont{J.}~\bibnamefont{Dominy}}, \bibnamefont{and}
  \bibinfo{author}{\bibfnamefont{H.}~\bibnamefont{Rabitz}},
  \bibinfo{journal}{Physical Review A (Atomic, Molecular, and Optical Physics)}
  \textbf{\bibinfo{volume}{79}}, \bibinfo{eid}{013422}
  (pages~\bibinfo{numpages}{16}) (\bibinfo{year}{2009}),
  \urlprefix\url{http://link.aps.org/abstract/PRA/v79/e013422}.

\bibitem[{\citenamefont{Wu et~al.}(2008{\natexlab{b}})\citenamefont{Wu,
  Chakrabarti, and Rabitz}}]{WuRaj2008}
\bibinfo{author}{\bibfnamefont{R.}~\bibnamefont{Wu}},
  \bibinfo{author}{\bibfnamefont{R.}~\bibnamefont{Chakrabarti}},
  \bibnamefont{and} \bibinfo{author}{\bibfnamefont{H.}~\bibnamefont{Rabitz}},
  \bibinfo{journal}{Physical Review A} \textbf{\bibinfo{volume}{77}},
  \bibinfo{eid}{052303} (pages~\bibinfo{numpages}{13})
  (\bibinfo{year}{2008}{\natexlab{b}}).

\bibitem[{\citenamefont{Ramakrishna et~al.}(1995)\citenamefont{Ramakrishna,
  Salapaka, Dahleh, Rabitz, and Pierce}}]{RamSal1995}
\bibinfo{author}{\bibfnamefont{V.}~\bibnamefont{Ramakrishna}},
  \bibinfo{author}{\bibfnamefont{M.~V.} \bibnamefont{Salapaka}},
  \bibinfo{author}{\bibfnamefont{M.}~\bibnamefont{Dahleh}},
  \bibinfo{author}{\bibfnamefont{H.}~\bibnamefont{Rabitz}}, \bibnamefont{and}
  \bibinfo{author}{\bibfnamefont{A.}~\bibnamefont{Pierce}},
  \bibinfo{journal}{Phy. Rev. A} \textbf{\bibinfo{volume}{51}},
  \bibinfo{pages}{960} (\bibinfo{year}{1995}).

\bibitem[{\citenamefont{Wu et~al.}(2006)\citenamefont{Wu, Tarn, and
  Li}}]{WuTarn2006}
\bibinfo{author}{\bibfnamefont{R.-B.} \bibnamefont{Wu}},
  \bibinfo{author}{\bibfnamefont{T.-J.} \bibnamefont{Tarn}}, \bibnamefont{and}
  \bibinfo{author}{\bibfnamefont{C.-W.} \bibnamefont{Li}},
  \bibinfo{journal}{Phys. Rev. A} \textbf{\bibinfo{volume}{73}},
  \bibinfo{pages}{012719} (\bibinfo{year}{2006}).

\bibitem[{\citenamefont{Boothby}(1975)}]{Boothby1975}
\bibinfo{author}{\bibfnamefont{W.}~\bibnamefont{Boothby}}, \bibinfo{journal}{J.
  Diff. Equa.} \textbf{\bibinfo{volume}{17}}, \bibinfo{pages}{296}
  (\bibinfo{year}{1975}).

\bibitem[{\citenamefont{Serovaiskii}(2004)}]{Sero2004}
\bibinfo{author}{\bibfnamefont{S.}~\bibnamefont{Serovaiskii}},
  \emph{\bibinfo{title}{Counterexamples in optimal control theory}}
  (\bibinfo{publisher}{Brill Academic Publishers}, \bibinfo{address}{The
  Netherlands}, \bibinfo{year}{2004}).

\bibitem[{\citenamefont{Boscain and Charlot}(2004)}]{BosCha2004}
\bibinfo{author}{\bibfnamefont{U.}~\bibnamefont{Boscain}} \bibnamefont{and}
  \bibinfo{author}{\bibfnamefont{G.}~\bibnamefont{Charlot}},
  \bibinfo{journal}{ESAIM: Control, Optimisation and Calculus of Variations
  (COCV)} \textbf{\bibinfo{volume}{10}}, \bibinfo{pages}{593}
  (\bibinfo{year}{2004}).

\bibitem[{\citenamefont{D'Alessandro}(2001)}]{Daless2001}
\bibinfo{author}{\bibfnamefont{D.}~\bibnamefont{D'Alessandro}},
  \bibinfo{journal}{IEEE. Trans. on Automatic Control}
  \textbf{\bibinfo{volume}{46}}, \bibinfo{pages}{866} (\bibinfo{year}{2001}).

\bibitem[{\citenamefont{Bonnard and Chyba}(2003)}]{BonChy2003}
\bibinfo{author}{\bibfnamefont{B.}~\bibnamefont{Bonnard}} \bibnamefont{and}
  \bibinfo{author}{\bibfnamefont{M.}~\bibnamefont{Chyba}},
  \emph{\bibinfo{title}{Singular trajectories and their role in control
  theory}}, vol.~\bibinfo{volume}{40} of \emph{\bibinfo{series}{Mathematiques
  and Applications}} (\bibinfo{publisher}{Springer}, \bibinfo{address}{Berlin},
  \bibinfo{year}{2003}).

\bibitem[{\citenamefont{Dominy and Rabitz}(2008)}]{Dominy2008}
\bibinfo{author}{\bibfnamefont{J.}~\bibnamefont{Dominy}} \bibnamefont{and}
  \bibinfo{author}{\bibfnamefont{H.}~\bibnamefont{Rabitz}},
  \bibinfo{journal}{Journal of Physics A: Mathematical and Theoretical}
  \textbf{\bibinfo{volume}{41}}, \bibinfo{pages}{205305 (21pp)}
  (\bibinfo{year}{2008}).

\bibitem[{\citenamefont{Stengel}(1994)}]{Stengel1994}
\bibinfo{author}{\bibfnamefont{R.}~\bibnamefont{Stengel}},
  \emph{\bibinfo{title}{Optimal control and estimation}}
  (\bibinfo{publisher}{Dover Publications}, \bibinfo{address}{New York},
  \bibinfo{year}{1994}).

\bibitem[{\citenamefont{Zhu and Rabitz}(1998)}]{ZhuRab1998}
\bibinfo{author}{\bibfnamefont{W.}~\bibnamefont{Zhu}} \bibnamefont{and}
  \bibinfo{author}{\bibfnamefont{H.}~\bibnamefont{Rabitz}},
  \bibinfo{journal}{The Journal of Chemical Physics}
  \textbf{\bibinfo{volume}{109}}, \bibinfo{pages}{385} (\bibinfo{year}{1998}).

\bibitem[{\citenamefont{Chitour et~al.}(2008)\citenamefont{Chitour, Jean, and
  Trelat}}]{Chitour2008}
\bibinfo{author}{\bibfnamefont{Y.}~\bibnamefont{Chitour}},
  \bibinfo{author}{\bibfnamefont{F.}~\bibnamefont{Jean}}, \bibnamefont{and}
  \bibinfo{author}{\bibfnamefont{E.}~\bibnamefont{Trelat}},
  \bibinfo{journal}{SIAM J. Control Optim.} \textbf{\bibinfo{volume}{47}},
  \bibinfo{pages}{1078} (\bibinfo{year}{2008}).

\end{thebibliography}

\end{document}